\documentclass[12pt,prl,aps,superscriptaddress]{revtex4}
\usepackage{graphicx}
\newcommand{\beq}[2]{\begin{equation}#1\label{#2}\end{equation}}
\newcommand{\ceq}[1]{(\ref{#1})}

\newfont{\mbld}{cmbx10 scaled 800}
\newfont{\cab}{cmsy10 scaled 1200}
\newfont{\scab}{cmsy10 scaled 1000}
\newfont{\bcall}{cmbsy10 scaled 1200}

\begin{document}
\title{The probability distribution of the average relative distance
  between two points in a dynamical chain}
\author{Franco Ferrari}
\email{ferrari@univ.szczecin.pl}
\author{Jaros{\l}aw Paturej}\email{jpaturej@univ.szczecin.pl}
\affiliation{Institute of Physics and CASA*, University of Szczecin,
  ul. Wielkopolska 15, 70-451 Szczecin, Poland}
\author{Thomas A. Vilgis} 
\email{vilgis@mpip-mainz.mpg.de}
\affiliation{Max Planck Institute for Polymer Research, 10
  Ackermannweg, 55128 Mainz, Germany}
\author{Tomasz Wydro}
\email{wydro@lpm.u-nancy.fr}
\affiliation{Laboratoire de Physique des Materiaux
UMR CNRS 7556, 
Universite Henri Poincare Nancy 1, B.P. 239 
F-54506 Vandoeuvre-les-Nancy Cedex, France}

\begin{abstract}
Subject of this letter is
the dynamics of a chain obtained performing
the continuous limit of a system of links
and beads.
In particular, the 
probability distribution of
the relative position between two points of
the chain averaged over a given
interval of time is computed.
The physical meaning of the obtained result is investigated in the
limiting case of a stiff chain.
\end{abstract}
\maketitle
\section{Introduction}\label{sec:intro}
We study in this letter the dynamics of a chain
obtained performing the continuous limit of a system of links and
beads.
This kind of problems has been addressed in the seminal paper of
Edwards and Goodyear \cite{EdwGoo} using an approach based on the
Langevin equation. 
Related papers dedicated to the statistical mechanics of a freely jointed
chain in the 
continuous limit are for example \cite{kraemers,grosberg,mazars, a-e}.
In Ref. \cite{FEPAVI1} it has been shown that it is possible to
investigate the dynamics of such a chain in a path integral
framework. The model which describes the fluctuations of the chain is
a generalization of the nonlinear sigma model \cite{nlsigma} called
the generalized 
nonlinear sigma model or simply GNLSM.
In \cite{FEPAVI1} it has also been discussed the relation between the
GNLSM and the Rouse model \cite{rouse,doiedwards}.
Applications of the GNLSM have been the subject of
Ref.~\cite{FEPAVI2}, in which the dynamic structure factor of the
chain has been computed in the semiclassical approximation.

The goal of the present work is to derive the probability distribution
$Z(\mathbf r_{12})$ which measures the probability that in a given
interval of time $\Delta t$ the average distance between two points of
the chain $P_1$ and $P_2$ is $\mathbf r_{12}$.
With respect to Ref.~\cite{FEPAVI2}, we use  to perform the calculation
a totally different 
approximation, which linearizes the
GNLSM. Moreover, a background 
field method is adopted, in which the effects of the thermal
fluctuations are
considered in the background of a fixed chain configuration 
chosen among the classical solutions
of the linearized equations of motion.
We find out that classical configurations are particularly stable
against the changes due to fluctuations.
The latter become relevant
only if they act on the chain for a significant amount
of time.

The material presented in this letter is divided as follows.
The GNLSM is linearized exploiting a gaussian approximation of the
functional Dirac delta function. Next, the probability distribution $Z(\mathbf r_{12})$
is computed within this approximation.
The asymptotic form  of $Z(\mathbf r_{12})$, which is valid in
the case in which the chain is stiff, is derived.
Finally, a discussion of the obtained result is presented.
\section{The generalized nonlinear sigma model in the gaussian
  approximation}\label{sec2} 
Let us consider the partition function of the GNLSM:
\beq{
Z=\int{\cal D}\mathbf R(t,\sigma) e^{-\tilde
  c\int_0^{t_f}\int_0^Nd\sigma \dot\mathbf R^2(t,\sigma)}
\delta(|\mathbf R'|-\ell)
}{parfun}
with $\dot\mathbf R=\partial \mathbf R/\partial t$ and
$\mathbf R'=\partial \mathbf R/\partial \sigma$.
The boundary conditions at $t=0$ and $t=t_f$ of the field $\mathbf R$
are respectively given by $
\mathbf R(0,\sigma)=\mathbf R_0(\sigma)$
and $\mathbf
R(t_f,\sigma)=\mathbf R_f(\sigma)$, where
$\mathbf R_0(\sigma)$ and $\mathbf R_f(\sigma)$ represent given static
conformations of the chain.
The boundary conditions with respect to $\sigma$ are periodic:
$
\mathbf R(t,\sigma)=\mathbf R(t,\sigma+N)$.
It was shown in Refs. \cite{FEPAVI1} and \cite{FEPAVI2} that the above
partition function describes the dynamics of a closed chain that is
the continuous version of a freely jointed chain consisting of links
and beads. The constant $\tilde c$ appearing in Eq.~\ceq{parfun} is
given by:
\beq{
\tilde c= c\ell\qquad\mbox{with}\qquad c=\frac{M}{4k_BT\tau L}
}{ctildedef}
Here $k_B$ denotes the Boltzmann constant, $T$ is the temperature and
$\tau$ is the relaxation time which characterizes the ratio of the
decay of the drift velocity of the beads.
$M$ and $L$ represent the total mass and the total length of the chain
respectively. 

Let us note that in Eq. \ceq{parfun} the trajectory of the chain
has been parametrized with the help of the dimensionless parameter
$\sigma$, which is related to the arc--length used in
Refs. \cite{FEPAVI1} and \cite{FEPAVI2} by the relation $s=\ell
\sigma$.
The scale of
length $\ell$ introduced with this parametrization is connected to
the
quantities $L$ and $N$
by the identity $N=\frac{L}{\ell}$. 
We remark also that, with
respect to Refs.~\cite{FEPAVI1} and \cite{FEPAVI2}, the constraint
$\mathbf R'^2=\ell^2$ has been replaced by the constraint $|\mathbf
R'|=\ell$ exploiting the property of the functional Dirac delta
function
$\delta(\mathbf R'^2-\ell^2 )=\delta( |\mathbf R'|-\ell) $.
A proof of this equation can be found in Ref.~\cite{FEPAVI1}.

In order to deal with the delta function in Eq.~\ceq{parfun} we use
the following gaussian approximation \cite{FEPAVI1}:
\beq{
\delta(|\mathbf R'|-\ell)\sim
\exp\left(\int_0^{t_f}dt\int_0^Nd\sigma\frac\nu2
\mathbf R^{\prime\, 2}\right)
}{deltaapproxtwo}
which is valid when the parameter $\nu$ is large, while $\ell$ is
small.
 As a consequence, the partition function $Z$ becomes:
\beq{
Z=\int{\cal D}\mathbf
R\exp\left[-\int_0^{t_f}dt\int_0^Nd\sigma 
\left(
\tilde c\dot\mathbf R^2+\frac \nu2\mathbf R^{\prime\, 2}
\right)\right]
}{parfunapp}
After the approximation \ceq{deltaapproxtwo} the chain may be regarded
as a gaussian chain consisting in a set of $N$  segments of
average length $\ell$. 

\section{The probability distribution of the average position between
  two points of the chain}

At this point we pick up  two points $P_1$ and
$P_2$ of the chain, for 
instance $\mathbf R(t,\sigma_1)$ and $\mathbf R(t,\sigma_2)$.
We wish to study how the relative position between these two points
changes due to  thermal fluctuations.
To this purpose, we introduce the following probability distribution:
\begin{eqnarray}
Z(\mathbf r_{12})&=&\int {\cal D} \mathbf R
\exp\left[
-\int_0^{t_f}dt\int_0^Nd\sigma\left(
\tilde c\dot\mathbf R^2+\frac\nu 2\mathbf R^{\prime\,2}
\right)
\right]\nonumber\\
&\times&\delta\left(
\mathbf r_{12}-\int_{t_1}^{t_2}\frac{dt}{\Delta t}
(\mathbf R(t,\sigma_2)-\mathbf R(t,\sigma_1))
\right)\label{zetar12}
\end{eqnarray}
We suppose that $0\le t_1\le t_2\le t_f$.
To evaluate the path integral in Eq.~\ceq{zetar12} it is convenient to
perform the splitting:
$
\mathbf R(t,\sigma)=\mathbf R_{cl}(t,\sigma)+\mathbf r(t,\sigma)$.
Here $\mathbf R_{cl}(t,\sigma)$ is a solution of the classical
equations of motion
$
\left[
\tilde c\frac{\partial^2}{\partial t^2}
+\frac\nu2\frac{\partial^2}{\partial \sigma^2}
\right]\mathbf R_{cl}=0$.
The fluctuations around the classical background $\mathbf
R_{cl}(t,\sigma)$ are taken into account by $\mathbf r(t,\sigma)$.
The boundary conditions for $\mathbf R_{cl}(t,\sigma)$ 
at the initial and final instants
are:
$
\mathbf R_{cl}(t_f,\sigma)=\mathbf R_f(\sigma)$
and
$\mathbf R_{cl}(0,\sigma)=\mathbf R_0(\sigma)$.
With respect to $\sigma$ periodic boundary conditions are assumed:
$
\mathbf R_{cl}(t_f,\sigma)=\mathbf R_{cl}(t,\sigma+N)$.
The 
solution of the classical equations of motion is complicated due
the presence of the non-trivial boundary conditions and will not be
reported
here. It can be found in standard books of mathematical methods in
physics, like for instance \cite{morsfesh}.
The only important information related to the background
fields that is needed in the probability distribution
$Z(\mathbf r_{cl})$
is the average relative
position $\mathbf r_{cl}$ of the points $P_1$ and $P_2$ 
for a given classical 
conformation $\mathbf R_{cl}(t,\sigma)$:
\beq{
\mathbf r_{cl}=\int_{t_1}^{t_2}\frac{dt}{\Delta t}\left(
\mathbf R_{cl}(t,\sigma_2)-\mathbf R_{cl}(t,\sigma_1)
\right)
}{smarcldef}
For the fluctuations $\mathbf r(t,\sigma)$ we choose Dirichlet boundary
conditions in time $
\mathbf r(t_f,0)=0$, $\mathbf r(0,\sigma)=0$.
The boundary conditions with respect to $\sigma$ are instead periodic:
$
\mathbf r(t,\sigma)=\mathbf r(t,\sigma+N)$.
After a few calculations, it is possible to rewrite $Z(\mathbf r_{12})$
in the form:
\beq{
Z(\mathbf r_{12})=\int_{-\infty}^{\infty}d^3\mathbf k e^{i\mathbf
  k\cdot(\mathbf r_{12}-\mathbf r_{cl}
)}e^{-A_{cl}}Z(\mathbf k)
}{zr12castone}
where $\mathbf r_{cl}$ and
$
A_{cl}=\int_0^{t_f}dt\int_0^Nd\sigma\left(
\tilde c\dot\mathbf R^2_{cl}+\frac{\nu}2\mathbf R^{\prime\,2}_{cl}
\right)$
take into account the  contribution of the classical
background,
while the fluctuations appear in  $Z(\mathbf k)$:
\beq{
Z(\mathbf k)=\int{\cal D}\mathbf r\exp\left[
-\int_0^{t_f}dt\int_0^Nd\sigma\left(
\tilde c\dot\mathbf r^2+\frac\nu2\mathbf r^{\prime\,2}+i\mathbf
J\cdot\mathbf r
\right)
\right]
}{zr12zk}
In Eq.~\ceq{zr12zk} 
the external current $\mathbf J$ is given by:
\beq{\mathbf J(t,\sigma)=\frac{\mathbf k}{\Delta t}
\left(
\delta(\sigma-\sigma_2)-\delta(\sigma-\sigma_1)
\right)
\theta(t_2-t)\theta(t-t_1)
}
{extcurdef}
This current has been introduced to provide a convenient way of
rewriting the integral 
$\int_{t_1}^{t_2}\frac{dt}{\Delta t}(\mathbf r(t,\sigma_2)-\mathbf
r(t,\sigma_1))$.

Let's now concentrate on the computation of the partition function of
the fluctuations $Z(\mathbf k)$.
The gaussian integration over the fields $\mathbf r(t,\sigma)$ may be
easily carried out and gives as a result:
\beq{
Z(\mathbf k)=Ce^{W(\mathbf k)}
}{ztildek}
where $C$ is a constant and
\beq{
W(\mathbf k)=\exp\left[
\frac 12\int_0^{t_f}dtdt'\int_0^Nd\sigma d\sigma'
G(t,\sigma;t',\sigma')\mathbf J(t,\sigma)\cdot \mathbf J(t',\sigma')
\right]
}{wtildek}
In the above equation $G(t,\sigma;t',\sigma')$ denotes the Green
function satisfying the equation:
\beq{
\left[
\tilde c\frac{\partial^2}{\partial
  t^2}+\frac{\nu}{2}\frac{\partial^2}{\partial \sigma^2}
\right]G(t,\sigma;t',\sigma')=\delta(t-t')\delta(\sigma-\sigma')
}{grefundefequ}
According to our settings, 
 we choose for
$G(t,\sigma;t',\sigma')$ Dirichlet boundary conditions at the instants $t=t_f$ and $t=0$
and periodic boundary conditions in $\sigma$ and $\sigma'$. 
To solve Eq.~\ceq{grefundefequ}, we decompose $G(t,\sigma;t',\sigma')$
in Fourier series as follows:
\beq{
G(t,\sigma;t',\sigma')=\sum_{n=-\infty}^{+\infty}g_n(t,t')e^{-2\pi
  i\frac\ell L(\sigma-\sigma')n}
}{gexp}
Substituting also the Fourier expansion of the periodic Dirac delta
function
$\delta(\sigma-\sigma')=\sum_{n=-\infty}^{+\infty} \frac\ell L
e^{-2\pi
  i\frac\ell L(\sigma-\sigma')n}
$ in Eq.~\ceq{grefundefequ} and solving for $g_n(t,t')$, we obtain
for $n=0$:
\beq{g_0(t,t')
=\frac 1{Lc}\theta(t'-t)t\frac{t'-t_f}{t_f}+
\frac1{Lc}\theta(t-t')t'\frac{t-t_f}{t_f}
}{gzttp}
and for $n\ne 0$:
\begin{eqnarray}
g_n(t,t')&=&A_n\theta(t-t')\sinh\beta_nt'\sinh\beta_n(t-t_f)\nonumber\\
&+&A_n\theta(t'-t)\sinh\beta_nt\sinh\beta_n(t_f-t')\label{gnttp}
\end{eqnarray}
with
\beq{
\beta_n=\frac{|n|\pi}{L}\sqrt{\frac{2\nu \ell}{c}}
}{betandef}
\beq{
A_n=-\frac{1}{\sqrt{2\nu\ell c}}\,\,
\frac{1}{|n|\pi\sinh\beta_nt_f}
}{Andef}
In Eqs.~\ceq{gzttp} and \ceq{gnttp} $\theta(t)$ denotes the Heaviside
$\theta-$function $\theta(t)=1$ for $t\ge 0$ and $\theta(t)=0$ for
$t<0$.
To complete the calculation of $Z(\mathbf k)$ in Eq.~\ceq{ztildek} we
rewrite $W(\mathbf k)$ as follows:
\beq{
W(\mathbf k)=\frac
12\sum_{n=-\infty}^{+\infty}\int_0^{t_f}dtdt'\int_0^Nd\sigma d\sigma'
g_n(t,t')e^{-2\pi i \frac \ell L (\sigma-\sigma')n}\mathbf
J(t,\sigma)\cdot \mathbf J(t',\sigma')
}{wkbar}
It is easy to show that the only non-zero contributions to the above
integral are those for which $n\ne 0$.
After integrating over the variables $\sigma,\sigma'$ and over one of
the time variables,
we obtain:
\begin{eqnarray}
W(\mathbf k)&=&\frac{2\mathbf k^2}{(\Delta t)^2}\sum_{n\ne 0}
\frac {A_n}{\beta_n}
\left[
1-\cos 2\pi n\left(\frac {\ell_2-\ell_1}{L}\right)
\right]\nonumber\\
&\times&\int_{t_1}^{t_2}dt\sinh\beta_n(t_f-t)(\cosh\beta_nt-
\cosh\beta_nt_1) \label{wkbartwo}
\end{eqnarray}
where we have put
$
\ell_2=\ell\sigma_2$ and $\ell_1=\ell\sigma_1$.
Remembering that the coefficients $A_n$ defined in Eq.~\ceq{Andef} are all
strictly negative, it is easy to realize that $W(\mathbf k)$
is
negative too.
This property of $W(\mathbf k)$ will be necessary in order to perform
the remaining integration over $\mathbf k$ in the probability
distribution  $Z(\mathbf r_{12})$ of Eq.~\ceq{zr12castone}.

A last integration over $dt$ in Eq.~\ceq{wkbartwo} delivers the
following result:
\begin{eqnarray}
W(\mathbf k)&=&\frac{2\mathbf k^2}{(\Delta t)^2}
\sum_{n\ne 0}
\frac{A_n}{\beta_n}
\left[
1-\cos 2\pi n\left(
\frac{\ell_2-\ell_1}{L}
\right)
\right]\left[
\frac 12(t_2-t_1)\sinh\beta_nt_f-\frac{\cosh\beta_nt_f}{2\beta_n}
\right.\nonumber\\
&-&\frac 1{4\beta_n}(\cosh\beta_n(t_f-2t_2)
+\cosh\beta_n(t_f-2t_1))\label{wkbarexact} \\
&+&\left.
\frac{1}{2\beta_n}(\cosh\beta_n(t_f+t_1-t_2)+\cosh\beta_n(t_f-(t_1+t_2)))
\right]\nonumber
\end{eqnarray}
Substituting Eq.~\ceq{wkbarexact} back in Eq.~\ceq{ztildek} we get an
exact expression for $Z(\mathbf
k)$.
To obtain the probability distribution $Z(\mathbf
r_{12})$ of Eq.~\ceq{zr12castone} in closed form a gaussian integration
over $\mathbf k$ is sufficient.

\section{Calculation of the probability distribution $Z(\mathbf
  r_{12})$ in the limit of a stiff chain}

To simplify our calculations of the previous Section, we compute the
infinite sum 
in Eq.~\ceq{wkbarexact} using an approximation.
First of all, we give a physical meaning to the parameter $\nu$ by
putting:
\beq{\nu=\frac{\alpha}{k_BT\tau\ell}}{nuphysint}
After performing the above substitution in 
the partition function of
Eq.~\ceq{parfunapp}, the
term proportional to $\nu$ appearing in the action
of the fields $\mathbf R$ becomes exactly the
term introduced in the
GNLSM in Ref.~\cite{FEPAVI2} in order to take into account
the bending energy of the chain.
This connection with the bending energy is confirmed by the fact
that $\alpha$ has dimensions of an energy per unit of
length,
i.~e. $[\alpha]=[\mbox{energy}]\cdot[\mbox{length}]^{-1}$.

As it was shown in Ref.~\cite{FEPAVI2},
large values of $\alpha$ correspond  
to a stiff chain.
Indeed, it is possible to check from Eq.~\ceq{zr12zk} that in this
case the corrections to the tangent vectors $\mathbf R_{cl}'$ coming
from the fluctuations $\mathbf r'$ are strongly suppressed in
$Z(\mathbf k)$. 
In the following, we will work in the limit of a stiff chain, i.~e.:
\beq{\alpha>>1}{stiffchain}
implies that also
From Eqs.~\ceq{nuphysint} and \ceq{stiffchain} it turns out that the
quantity
$\nu\ell=\frac {\alpha}{k_BT\tau}$
appearing in the
coefficients $\beta_n$ and $A_n$ (see Eqs.~\ceq{betandef} and \ceq{Andef})
is very large. 
As a consequence, within the approximation
\ceq{stiffchain}
the coefficients $\beta_n$ are large, while the
coefficients $A_n$ are exponentially small.
Taking into account these facts, we may write the following asymptotic
expression of
$W(\mathbf k)$:
\beq{
W(\mathbf k)=\frac{2\mathbf k^2}{(\Delta t)^2}\frac{L}{2\pi^2\nu\ell}
\left[
-A\frac{(t_2-t_1)}{2}+B\frac{L}{\pi}\sqrt{\frac{c}{2\nu\ell}}
\right]+{\cal O}\left(
e^{-\frac{2\pi}{L}\sqrt{\frac{2\nu\ell}{c}}(t_2-t_1)}
\right)
}{wkbarapprox}
In Eq.~\ceq{wkbarapprox} we have introduced the convenient notation:
\begin{eqnarray}
A&=&\sum_{n\ne 0}\frac 1{n^2}\left[
1-\cos 2\pi n\left(\frac{\ell_2-\ell_1}{L}\right)
\right]\\
B&=&\frac 12\sum_{n\ne 0}\frac 1{|n|^3}\left[
1-\cos 2\pi n\left(
\frac{\ell_2-\ell_1}{L}
\right)
\right]
\end{eqnarray}
It is possible to check that the other contributions to $W(\mathbf
k)$, expressed in Eq.~\ceq{wkbarapprox} with the symbol
${\cal O}\left(
e^{-\frac{2\pi}{L}\sqrt{\frac{2\nu\ell}{c}}(t_2-t_1)}
\right)$, decay at least as fast as
$e^{-\frac{2\pi}{L}\sqrt{\frac{2\nu\ell}{c}}(t_2-t_1)}$.
These terms become negligibly small when $\nu\ell$ is large, provided
of course that:
\beq{
  t_2-t_1>>\frac12\frac L\pi\sqrt{\frac{c}{2\nu\ell}}
}{condneg}
Substituting Eq.~\ceq{wkbarapprox} in Eq.~\ceq{ztildek}, we obtain for
$Z(\mathbf k)$ the
approximate expression: 
\beq{
Z(\mathbf k)=Ce^{-\frac{\kappa\mathbf k^2}{2}}
}{zkfinal}
where
\beq{
\kappa=\frac 4{(\Delta t)^2}\frac{L}{\pi^22\nu\ell}\left[
A\frac{t_2-t_1}{2}-B\frac L\pi\sqrt{\frac{c}{2\nu\ell}}
\right]
}{alphapardef}
The condition \ceq{condneg} guarantees that $\kappa>0$.
This is what is needed to compute the integral over $\mathbf k$ in
Eq.~\ceq{zr12castone}.
After a few calculations we arrive at the final result:
\beq{
Z(\mathbf r_{12})=Ce^{-A_{cl}} \left(
\frac{2\pi}{\kappa}
\right)^{\frac 32}\exp\left[
-\frac 12\frac{(\mathbf r_{12}-\mathbf r_{cl})^2}{\kappa}
\right]
}{probdistfina}
Eq.~\ceq{probdistfina} has a straightforward interpretation.
The quantity $\kappa$  is
inversely proportional to the product $\nu\ell$, which is supposed to
be large. 
Thus $\kappa$ is very small. As a consequence, 
Eq.~\ceq{probdistfina} implies that
the relative positions
of the two points $P_1$ and $P_2$ exhibits a sharp peak around the
classical value, 
i.~e. when $\mathbf r_{12}=\mathbf r_{cl}$.
From Eq.~\ceq{probdistfina} it turns also out that the changes
of the classical background configuration
$\mathbf R_{cl}(t,\sigma)$ due to the fluctuations $\mathbf
r(t,\sigma)$ are relatively small. This is  due to the fact that there
is no contribution to the parameter $\kappa$ which is of zeroth order
in $\nu\ell$. Potentially, a zeroth order contribution could come from
the term proportional to $g_0(t,t')$ 
in Eq.~\ceq{wkbar}, but we have seen that this term vanishes
identically.
This suggests that the effects of thermal fluctuations are weak at short
time-scales. Of course, with the increasing of the time interval
$\Delta t=t_2-t_1$ over which the average of the relative position of
$P_1$ and $P_2$ is measured, the fluctuations have more and more time
to act on the chain and their influence becomes more and more important.
This intuitive prediction is confirmed by Eq.~\ceq{probdistfina}.
Indeed, the coefficient $\kappa$ grows linearly with $\Delta t$ and,
for larger values of $\kappa$, the peak around the classical point
$\mathbf r_{12}=\mathbf r_{cl}$ in the probability distribution
$Z(\mathbf r_{12})$ becomes less sharp as expected.
\section{Conclusions}
In this work  the GNLSM of Refs.~\cite{FEPAVI1} and
\cite{FEPAVI2} has been applied to compute the distribution function
$Z(\mathbf r_{12})$ of the relative position between two points of a
chain with rigid constraints. The calculation has been performed using
the approximation 
\ceq{deltaapproxtwo} of the functional Dirac delta function which is
needed to impose the  constraints. After this approximation, the
finest details of the chain are lost, a fact that has already been
noted
in the case of a static chain \cite{EdwGoo}.

A closed form of 
the probability distribution \ceq{probdistfina}
may be obtained
starting from the exact calculation of the
contribution of the fluctuations to $Z(\mathbf r_{12})$ given in
Eq.~\ceq{wkbarexact}. The final formula for $Z(\mathbf r_{12})$
computed in this way is however complicated.
For this reason
the physical meaning of  $Z(\mathbf r_{12})$
has been investigated 
at the end of the previous Section
in the case in which 
the energy needed for bending the
chain is large. In this approximation the
contributions of the 
fluctuations which decay exponentially when $\alpha$
becomes large are neglected.

Concluding, it would be interesting to explore the connections between
the GNLSM and other models of the dynamics of a chain, in which
instead of rigid constraints the addition of potentials is used in
order to prevent the breaking of the chain \cite{FEMIROVI}.
\section{Acknowledgements}
This work has been financed
by the Polish Ministry of Science and Higher Education, scientific
project N202 156 
31/2933.  
F. Ferrari gratefully acknowledges also the support of the action
COST~P12 financed by the European Union and the hospitality of
C. Schick at the University of Rostock.
The authors  would like to thank
V. G. Rostiashvili for fruitful discussions.

\end{document}